\documentclass[english,prl,twocolumn]{revtex4}
\usepackage[T1]{fontenc}
\usepackage[latin9]{inputenc}
\usepackage{graphicx}
\usepackage{esint}

\makeatletter
\@ifundefined{textcolor}{}
{%
 \definecolor{BLACK}{gray}{0}
 \definecolor{WHITE}{gray}{1}
 \definecolor{RED}{rgb}{1,0,0}
 \definecolor{GREEN}{rgb}{0,1,0}
 \definecolor{BLUE}{rgb}{0,0,1}
 \definecolor{CYAN}{cmyk}{1,0,0,0}
 \definecolor{MAGENTA}{cmyk}{0,1,0,0}
 \definecolor{YELLOW}{cmyk}{0,0,1,0}
}

\makeatother

\usepackage{babel}
\begin{document}

\title{Noise driven emergence of cooperative behavior}

\author{Bahram Houchmandzadeh}

\address{Univ. Grenoble 1/CNRS, LIPhy UMR5588, Grenoble F-38401, France.}
\begin{abstract}
Cooperative behaviors are defined as the production of common goods
benefitting all members of the community at the producer's cost. They
could seem to be in contradiction with natural selection, as non-cooperators
have an increased fitness compared to cooperators. Understanding the
emergence of cooperation has necessitated the development of concepts
and models (inclusive fitness, multilevel selection, ...) attributing
deterministic advantages to this behavior. In contrast to these models,
we show here that cooperative behaviors can emerge by taking into
account the stochastic nature of evolutionary dynamics : when cooperative
behaviors increase the carrying capacity of the habitat, they also
increase the genetic drift against non-cooperators. Using the Wright-Fisher
models of population genetics, we compute exactly this increased genetic
drift and its consequences on the fixation probability of both types
of individuals. This computation leads to a simple criterion: cooperative
behavior dominates when the relative increase in carrying capacity
of the habitat caused by cooperators is higher than the selection
pressure against them. This is a purely stochastic effect with no
deterministic interpretation. 
\end{abstract}
\maketitle

\section{Introduction.}

Cooperative behaviors can be defined as the production a common good
by an individual that benefits everybody in the community. Such behavior
has a cost in terms of fitness, as the producer devotes part of its
resources to this task. To early evolutionary biologists, cooperative
behaviors seemed to be in contradiction with natural selection\citep{Sober1999,Dugatkin2006}:
since selection acts on individuals, a non-cooperator has a higher
fitness than a cooperator and will always invade the community. Cooperative
behaviors however, specially in microbial world, are widespread. A
few examples of such behaviors are light production in \emph{Vibrio
fisheri}\citep{Visick2006}, siderophore production in \emph{Pseudomonas
aeroguinosa}\citep{West2003,Harrison2009}, stalk formation by \emph{Dictyostelium
discoidum}\citep{Kessin2001,Foster2004}, decreased virulence in many
pathogen-host systems\citep{Diggle2010}. All these cases are examples
of a \emph{production of common good} by an individual benefiting
every individual in the community. More generally, these behaviors
can be seen as particular cases of Niche Construction\citep{Odling-smee2013}.

Researchers have investigated the deterministic advantages that these
kinds of behaviors could confer on individuals. The major schools
along this line of investigation are inclusive fitness\citep{Hamilton1964a,Michod1982,Gardner2011}
and multilevel selection\citep{Lewontin1970,Wilson1983,Silva1999,Traulsen2006a}
and their associated variants\citep{Nowak2006}, although the relative
merits of these concepts are sometimes hotly debated\citep{Lehmann2007,Allen2013a}.
The aim of this article is not to discuss the relevance of these models,
which have been documented in a large number of books and articles.
The fact that cooperative behaviors are so widespread, however, behooves
us to search for simple mechanisms to explain their emergence. I intend
in this article to show that cooperative behaviors, by the simple
act of increasing the carrying capacity of the habitat, give an advantage
to cooperators. The origin of this advantage is not deterministic,
but has to be sought in the stochasticity of evolutionary dynamics. 

Evolution is an interplay between deterministic causes broadly called
fitness, and random events such as sampling between generations. An
advantageous mutant does not spread with certainty but has only a
greater probability of invading the community and of being fixed.
This probability, called the fixation probability, is the relevant
quantity to weight deterministic versus stochastic causes\citep{Patwa2008a}. 

Consider an asexual population of fixed size $N$, with two types
of individuals $A$ and $S$, where $S$ types have a constant excess
relative fitness $s$ compared to $A$. the deterministic differential
equation describing the variation of the proportion $x=N_{A}/N$ of
the $A$ type is\citep{Ewens2004}: 
\begin{equation}
dx/dt=-sx(1-x)\label{eq:determinstic.}
\end{equation}
and leads to the disappearance of $A$ individuals ($x\rightarrow0$). 

Going beyond the deterministic approach, one can solve the full stochastic
dynamics of such a model and extract the invasion capacity of these
two types, $i.e.$ the fixation probability $\pi_{1}^{i}$ of one
individual of type $i$ introduced into a population consisting entirely
of the other type. In the framework of the Wright-Fisher or Moran
model of population genetics, for a population of fixed size $N$,
in the small selection pressure limit $Ns\ll1$: 
\begin{eqnarray}
\pi_{1}^{A} & = & \frac{1}{N}-s\label{eq:pi1A}\\
\pi_{1}^{S} & = & \frac{1}{N}+s\label{eq:pi1S}
\end{eqnarray}
Therefore, if $s>0$ then $\pi_{1}^{S}>\pi_{1}^{A}$ and type $S$
individuals have a higher invasion capacity than type $A$ individuals.
In this case, the ratio of invasion capacities has the same information
content as the deterministic approach: both lead to the conclusion
that $s>0$ favors the $S$ type. This broad equivalence between these
two approaches has led researchers to investigate the existence of
\emph{deterministic} advantages that could favor the cooperators ($A$
individuals) against non-cooperators ($S$ individuals) even though
$s$, the \emph{bare} fitness of $S$ (or equivalently, the cost of
altruism to $A$) is positive. 

Fluctuations and random events can however be more subtle and alter
the equivalence between deterministic and stochastic modeling. In
particular, we can have $\pi_{1}^{S}<\pi_{1}^{A}$ \emph{even when}
$s>0$, without any hidden deterministic advantage. This is the case
of a cooperative behavior that increases the carrying capacity of
the habitat. Consider a system where the carrying capacity is a function
of the proportion of cooperators, varying between $N_{i}$ when only
$S$ type is present and $N_{f}$ where only $A$ type is present,
with $N_{i}<N_{f}$ (figure \ref{fig:SchemeVarN}). We suppose that
$S$ types have a constant excess relative fitness $s>0$. The deterministic
equation (\ref{eq:determinstic.}) does not change and will again
lead to the $A$'s extinction. In contrast to the deterministic computation
however, a back of the envelope estimation of the invasion capacity
of both types yields:
\begin{figure}
\begin{centering}
\includegraphics[width=0.8\columnwidth]{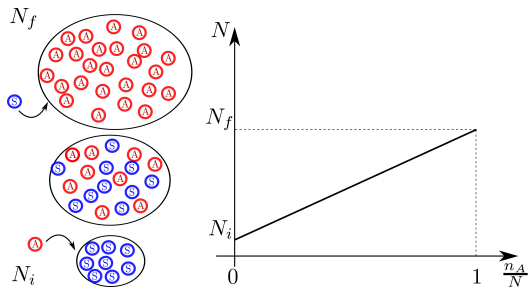}
\par\end{centering}

\caption{Scheme of a cooperative behavior where the carrying capacity $N$
of the habitat is an increasing function of the proportion $x$ of
type $A$ individuals : $N=N(x)$. For a habitat formed of only $S$
type individuals, $N(0)=N_{i}$. When only $A$ individuals are present,
$N(1)=N_{f}$, where $N_{f}>N_{i}$. The invasion capacity of each
type is defined as the fixation probability of one $i$ type introduced
into a community formed only of type $j$. \label{fig:SchemeVarN}}

\end{figure}
\begin{eqnarray}
\pi_{1}^{A} & = & \frac{1}{N_{i}}-s\label{eq:pi1A-1}\\
\pi_{1}^{S} & = & \frac{1}{N_{f}}+s\label{eq:pi1S-1}
\end{eqnarray}
We observe that we can have $\pi_{1}^{S}<\pi_{1}^{A}$ even though
$s>0$, if 
\[
2s<\frac{1}{N_{i}}-\frac{1}{N_{f}}
\]
or equivalently, on setting $\bar{N}=\sqrt{N_{i}N_{f}}$, if $2\bar{N}s<\Delta N/\bar{N}$:
if the \emph{selection pressure} against cooperators is smaller than
the relative variation in the carrying capacity due to cooperators,
then the latter type is favored and has a higher invasion capacity.
This is a purely stochastic effect with no deterministic counterpart
and is due to the fact that cooperators increase the genetic drift
of non-cooperators. 

We had previously shown the existence of this effect using a two dimensional
Markov chain approach of a generalized Moran model\citep{Houchmandzadeh2012a}.
This approach however was mathematically intricate and only approximate
solutions could be obtained at small selection pressure. The effect
however can be understood in a much simpler way using a classical
Wright-Fisher (WF) model of population genetics, which I develop in
the following sections, where very general results can be obtained.
The WF model is a well studied generic model of population genetics,
shown to be equivalent to many other models of population genetics\citep{Lambert2006}. 

The article is organized as follow. In the result section, the first
subsection is devoted to recalling the main results of the classical
WF model. In the second subsection, a simple system is considered
where the carrying capacity is a linear function of the proportion
of cooperators. An exact solution for the fixation probabilities is
then obtained and it is shown that cooperators can have a higher invasion
capacities than non-cooperators, even when the cost of cooperation
is always positive. The third subsection generalizes this concept
to arbitrary dependence of the carrying capacity on the proportion
of cooperators; a very simple and general criterion is then obtained
for cooperators to prevail. Finally, the extension to diploid populations
is considered in the following subsection. The conclusion section
put these results into perspective in respect to other models of the
emergence of cooperation.

\section{Results.}

\subsection{Preliminaries.}

I recall the main results of the classical Wright-Fisher model for
the sake of clarity of the following sections. In a community of fixed
size $N$, two types of asexual adult individuals $A$ and $S$ of
abundance $n_{A}$ and $n_{S}$ ($n_{A}+n_{S}=N)$ produce progeny.
This progeny is then sampled to form the next generation of adults.
The sampling process is biased toward the $S$ type which has an excess
relative fitness $s$ which we suppose to be small ($0<s\ll1$). 

The transition probability to have $k$ individuals of type $A$ in
the next generation $G_{i+1}$ when $n_{A}$ individuals were present
at generation $G_{i}$ is binomial \citep{Moran,Ewens2004}:
\[
P\left(k|n_{A}\right)={N \choose k}u^{k}(1-u)^{N-k}
\]
where 
\begin{eqnarray}
u & = & \frac{x}{x+(1+s)(1-x)}\label{eq:define:u}\\
 & = & x-sx(1-x)+{\cal O}(s^{2})
\end{eqnarray}
In the above expression, $x=n_{A}/N$ designates the proportion of
$A$ in $G_{i}$. The bias $s$ toward the selection of one type can
be due to the increase in mean number of progeny, the variability
in their production\citep{Gillespie1974,Lambert2006}, or any other
similar phenomena. 

The exact dynamics of the above stochastic process is not known, but
one can resort to the diffusion approximation \citep{Kimura1962,Ewens2004}
to compute various quantities of interest. This computation is based
on the change in the mean and variance of the proportion of $A$ types
in the next generation, which, to the first order in $s$ is: 
\begin{eqnarray}
a(x) & = & \left\langle y|x\right\rangle -x=-sx(1-x)\label{eq:drift}\\
b(x) & = & \frac{1}{2}\mbox{Var}(y|x)=\frac{1}{2N}x(1-x)\label{eq:variance}
\end{eqnarray}
where $y$ is the proportion of $A$ in the next generation; $\left\langle y|x\right\rangle $
designates the expectation of $y$ conditioned on $x$, the proportion
of $A$ in the present generation. The fixation probability $\pi(x)$
of the $A$ type present with proportion $x$ at the first generation
can be computed from the backward diffusion approximation of stochastic
dynamics\citep{Kimura1962} : 
\begin{equation}
a(x)\frac{d\pi(x)}{dx}+b(x)\frac{d^{2}\pi(x)}{dx^{2}}=0\label{eq:kimurafix}
\end{equation}
As $-a(x)/b(x)=2Ns$, the use of boundary conditions $\pi(0)=0$,
$\pi(1)=1$ leads to the well known Kimura solution: 
\begin{eqnarray}
\pi(x) & = & \frac{1-e^{2Nsx}}{1-e^{2Ns}}\label{eq:fixresult}\\
 & \approx & x-Nsx(1-x)\,\,\,\mbox{for}\,\,\, Ns\ll1
\end{eqnarray}
The invasion capacity of both types are readily obtained from the
above expression 
\[
\pi_{1}^{A}=\pi(1/N)\,\,\,;\,\,\,\pi_{1}^{S}=1-\pi(1-1/N)
\]
and are equal to expressions (\ref{eq:pi1A},\ref{eq:pi1S}) in the
small selection pressure limit $Ns\ll1$. The ratio of invasion capacities
reads 
\begin{equation}
R_{SA}=\frac{\pi_{1}^{S}}{\pi_{1}^{A}}=\frac{1+Ns}{1-Ns}\approx1+2Ns\label{eq:RSAfix}
\end{equation}
and $R_{SA}>1$ if $s>0$. Note that here we use $s$ as the relative
excess advantage of the $S$ type, or equivalently, the \emph{cost}
of $A$ type. Hence the change in the sign of $s$ in expression (\ref{eq:fixresult})
compared to similar expressions used in the literature. The reason
behind this choice is that in the following, $A$ will designates
the cooperators with a positive cost for cooperation.

\subsection{Variable size population.}

\begin{figure}
\begin{centering}
\includegraphics[width=0.8\columnwidth]{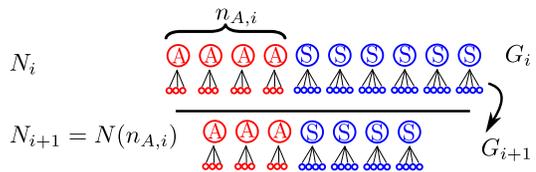}
\par\end{centering}

\caption{Scheme of sampling between generations in the present model : The
carrying capacity $N_{i+1}$ of generation $G_{i+1}$ is a function
of the number of cooperators $n_{A,i}$ present at generation $G_{i}$
: $N_{i+1}=N(n_{A})$. The selection process consists in selecting
$N_{i+1}$ individuals among the progeny at $G_{i}$. \label{fig:variablesampling}}
\end{figure}

Consider now a system in which the carrying capacity is not constant,
but is an increasing function of the number of cooperators $n_{A}$(figure
\ref{fig:SchemeVarN}). The stochastic behavior of such a system can
be modelled as follow: as in the fixed size habitat before, both $A$
and $S$ types at generation $G_{i}$ produce progeny ; however, the
carrying capacity $N_{i+1}$ of the next generation $G_{i+1}$ depends
on the number (or proportion) of cooperators in $G_{i}$. Hence $N_{i+1}=N(n_{A,i})$
individuals among the progeny are randomly selected to form the next
generation (figure \ref{fig:variablesampling}).
The probability of having $k$ individuals of type $A$ in generation
$G_{i+1}$, knowing that there were $n_{A,i}$ individuals at generation
$G_{i}$ is binomial :
\begin{equation}
P\left(k|n_{A,i}\right)={N_{i+1} \choose k}u^{k}(1-u)^{N_{i+1}-k}\label{eq:sampling}
\end{equation}
where $u$ has the same definition as in (\ref{eq:define:u}). We
can repeat all the arguments for the computation of relative change
in the mean and variance of the proportion $x=n_{A}/N$ of $A$ types,
keeping in mind that the only difference in the present model is that
$N=N(x)$ is no longer constant. In particular, the fixation probabilities
are given by the same backward diffusion equation 
\begin{equation}
a(x)\frac{d\pi(x)}{dx}+b(x)\frac{d^{2}\pi(x)}{dx^{2}}=0\label{eq:fixation2}
\end{equation}
where this time, 
\[
-\frac{a(x)}{b(x)}=2sN(x)
\]
For the sake of simplicity, in this subsection we suppose a linear
form for $N(x)$ (figure \ref{fig:SchemeVarN}): 
\begin{equation}
N(x)=(1-x)N_{i}+xN_{f}\label{eq:CC}
\end{equation}
where $N_{f}$ and $N_{i}$ ($N_{i}<N_{f}$) are the carrying capacities
of the habitat when composed only of \emph{A} types and $S$ type.
The next subsection generalizes the computation to arbitrary form
of $N(x)$. The differential equation (\ref{eq:fixation2}) can still
be easily solved. Let us express $N_{i}$ and $N_{f}$ in terms of
their mean and relative difference 
\begin{eqnarray}
\bar{N} & = & (N_{f}+N_{i})/2\label{eq:Nbar}\\
2\delta & = & (N_{f}-N_{i})/\bar{N}\label{eq:delta}
\end{eqnarray}
Then, setting 
\[
\psi(x)=2\bar{N}sx\left(1-\delta+\delta x\right)
\]
we have 
\begin{equation}
\frac{d\pi}{dx}=C\exp\left(\psi(x)\right)\label{eq:dpidx}
\end{equation}
Integrating once more and taking into account the boundary conditions
$\pi^{A}(0)=0$, $\pi^{A}(1)=1$, the solution can be written as 
\begin{equation}
\pi(x)=\frac{f(0)-f(x)}{f(0)-f(1)}\label{eq:general_fix}
\end{equation}
where 
\[
f(x)=\frac{e^{2\bar{N}s\left((1-\delta)x+\delta x^{2}\right)}}{\sqrt{2\bar{N}s\delta}}D\left(\sqrt{\frac{\bar{N}s}{2\delta}}(1-\delta+2\delta x)\right)
\]
and 
\[
D(z)=e^{-z^{2}}\int_{0}^{z}e^{u^{2}}du
\]
is the Dawson function. In the case of small selection pressure against
$A$ ($\bar{N}s\ll1$) and small relative change in the size of the
habitat $\delta\ll1$, expanding the fixation probability (\ref{eq:general_fix})
to the second order in $\bar{N}s$ and $\delta$, we find 
\begin{equation}
\pi(x)\approx x-\bar{N}sx(1-x)+\frac{\bar{N}s}{3}(\bar{N}s+\delta)x(1-x)(1-2x).\label{eq:fixsecondorder}
\end{equation}
Note that this limit could have been obtained by direct integration
of the power series expansion of $\exp\left(\psi(x)\right)$ in expression
(\ref{eq:dpidx}). Direct integration of power series allows more
complicated laws of carrying capacity $N(x)$ to be taken into account,
when exact solutions are no longer available. 

\begin{figure}
\begin{centering}
\includegraphics[width=1\columnwidth]{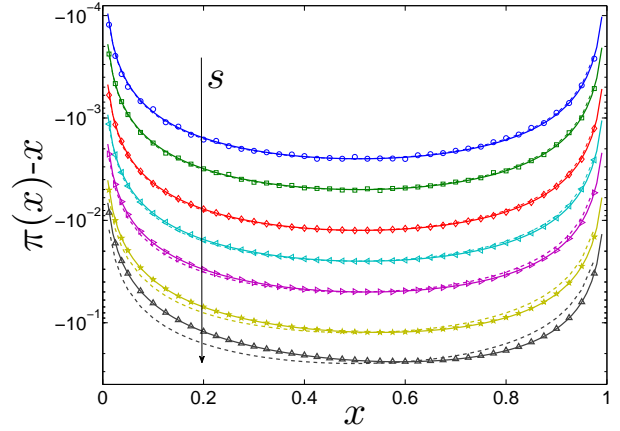}
\par\end{centering}

\caption{Numerical computation of $\pi(x)$ and its comparison with theoretical
values for $N_{i}=90$, $N_{f}=110$ and $s=(1,2,5,10,20,50,100)\times10^{-4}$.
The arrow indicates the direction of increasing $s$. For each value
of initial proportion of $A$ individuals $x\in[0,1]$, a random discrete
path is generated by determining the carrying capacity in the next
generation according to relation (\ref{eq:CC}) and then sampling
of the present generation to constitute the next generation according
to relation (\ref{eq:sampling}); the process is stopped by the loss
of either $A$ or $S$ individuals. $10^{8}$ such paths are generated
to compute, for each initial proportion $x$, its fixation probability
$\pi(x)$. Symbols : values obtained from numerical simulations ;
solid lines : theoretical values according to expression (\ref{eq:general_fix})
; dashed lines : small selection pressure approximation (eq. \ref{eq:fixsecondorder}).\label{fig:Numericalpi}}
\end{figure}

Figure \ref{fig:Numericalpi} shows the excellent agreement between
numerical simulations of the above stochastic process and the theoretical
predictions (eqs. \ref{eq:general_fix},\ref{eq:fixsecondorder}).

\begin{figure}
\begin{centering}
\includegraphics[width=1\columnwidth]{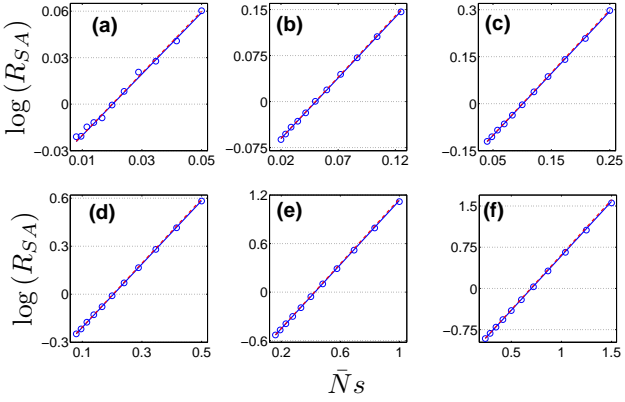}
\par\end{centering}

\caption{Logarithm of the ratio of invasion capacities $R_{SA}=\pi_{1}^{S}/\pi_{1}^{A}$
as a function of selection pressure $\bar{N}s$ for $\bar{N}=100$
and increasing value of $\delta=0.02,0.05,0.1,0.2,0.4,0.6$ (a-f).
Symbols : fixation probability ratio obtained from numerical simulations
as detailed in figure \ref{fig:Numericalpi}; solid lines : theoretical
predictions of fixation probabilities obtained from eq. (\ref{eq:general_fix})
; dashed lines : general solution given by relation (\ref{eq:RSAgeneral}).\label{fig:Ratio-of-invasion}}

\end{figure}

Let us now consider the case where $\delta$ and $\bar{N}s$ are of
the same order and compute the invasion capacity of each type by keeping
only the first order terms :
\begin{eqnarray*}
\pi_{1}^{A} & = & \pi\left(\frac{1}{N_{i}}\right)\approx\frac{1}{N_{i}}-\frac{\bar{N}}{N_{i}}s\\
\pi_{1}^{S} & = & 1-\pi\left(1-\frac{1}{N_{f}}\right)\approx\frac{1}{N_{f}}+\frac{\bar{N}}{N_{f}}s
\end{eqnarray*}
The ratio of invasion capacities now becomes:
\begin{eqnarray}
R_{SA} & = & \frac{\pi_{1}^{S}}{\pi_{1}^{A}}=\left(\frac{1-\delta}{1+\delta}\right)\left(\frac{1+\bar{N}s}{1-\bar{N}s}\right)\label{eq:Rapp1}\\
 & \approx & 1+2\bar{N}s-2\delta\label{eq:Rapp2}
\end{eqnarray}
We see that, contrary to the fixed population size case (eq. \ref{eq:RSAfix}),
we can have $R_{SA}<1$ even when $s>0$ ! The criterion for cooperators
to prevail is simply 
\begin{equation}
\delta>\bar{N}s\label{eq:criterium}
\end{equation}
Figure \ref{fig:Ratio-of-invasion} shows the excellent agreement
between $R_{SA}$ obtained from numerical results and the theoretical
prediction (eq. \ref{eq:general_fix}). Note that for large $\delta$,
the criterion (\ref{eq:criterium}) underestimates the advantage of
$A$ individuals, which can prevail at a higher cost of cooperation.
The general form of the criterion is given in the next subsection

Let us stress again that the $A$ type advantage is due purely to
the stochasticity of natural selection and has no deterministic  interpretation.
It is due to the interplay between genetic drift and deterministic
effect. The deterministic equation, neglecting fluctuations, is as
before 
\[
\frac{dx}{dt}=a(x)=-sx(1-x)
\]
and predicts the disappearance of $A$ individuals that have a negative
relative excess fitness.

\subsection{Solution for general form of $N(x)$.}

The conditions necessary for $R_{SA}<1$ can be obtained for any form
of $N(x)$ without computing $\pi(x)$ explicitly. First, note that
the first integral of the Kimura equation (\ref{eq:fixation2}) is
\begin{eqnarray*}
\log\left(\frac{\pi'(x)}{\pi'(0)}\right) & = & -\int_{0}^{x}\frac{a(u)}{b(u)}du\\
 & = & 2s\int_{0}^{x}N(u)du
\end{eqnarray*}
The invasion capacities can be expressed in terms of the function
$\pi'(x)$ : 
\begin{eqnarray*}
\pi_{1}^{A} & = & \pi(1/N_{i})=\pi'(0)(1/N_{i})+{\cal O}(1/N_{i}^{2})\\
\pi_{1}^{S} & = & 1-\pi(1-1/N_{f})=\pi'(1)(1/N_{f})+{\cal O}(1/N_{f}^{2})
\end{eqnarray*}
because of the boundary conditions $\pi(0)=0$ and $\pi(1)=1$. Therefore,
to the first order in $1/N$, the ratio of invasion capacities is:
\[
R_{SA}=\frac{\pi_{1}^{S}}{\pi_{1}^{A}}=\frac{\pi'(1)}{\pi'(0)}\frac{N_{i}}{N_{f}}
\]
and therefore
\begin{equation}
\log\left(R_{SA}\right)=2s\int_{0}^{1}N(u)du+\log\left(N_{i}/N_{f}\right)\label{eq:RSAgeneral}
\end{equation}
The condition for $A$ to prevail, $R_{SA}<1$, is then equivalent
to 
\begin{equation}
2s\int_{0}^{1}N(u)du<\log\left(N_{f}/N_{i}\right)\label{eq:nicecriterion}
\end{equation}
For the simple case investigated in the preceding subsection, in which
$N(x)=\bar{N}(1-\delta+2\delta x)$ the left hand side of relation
(\ref{eq:nicecriterion}) evaluates to $2\bar{N}s$ and the criterion
becomes 
\[
2\bar{N}s<\log\left(\frac{1+\delta}{1-\delta}\right)
\]
which reduces to expression (\ref{eq:criterium}) for small $\delta$.
The accuracy of this criterion, for the case of linear $N(x)$, is
shown in figure \ref{fig:Ratio-of-invasion}. The general criterion
(\ref{eq:nicecriterion}) is not limited to small selection pressure
or small variation in population size. The condition $N\gg1$ is still
necessary for the validity of the diffusion approximation\citep{Ethier1977}.

\subsection{Extension to diploid populations.}

The above results can be generalized to diploid, randomly mating populations
when cooperative behavior is caused by a single gene. Consider a diploid
population of size $N$ corresponding to $2N$ gametes. The fitness
of $(AA,AS,SS)$ individuals will be denoted $(1,1+s(1/2-\epsilon),1+s)$
where $s$, as before is the relative fitness value of the allele
$S$ and $\epsilon\in[-1/2,1/2]$ is the dominance of allele $A$
; $\epsilon=0$ corresponds to no dominance. As before, $x$ will
designate the frequency of allele $A$. Following the arguments of
the previous section, we can write the number of $A$ allele in the
next generation as 
\[
P\left(k|n_{A,i}\right)={2N_{i+1} \choose k}u^{k}(1-u)^{2N_{i+1}-k}
\]
where \citep{Ewens2004}
\begin{eqnarray*}
u & = & \frac{x^{2}+(1+s(1/2-\epsilon))x(1-x)}{x^{2}+2(1+s(1/2-\epsilon))x(1-x)+(1+s)(1-x)^{2}}\\
 & = & x-sx(1-x)(1+2\epsilon-4\epsilon x)+{\cal O}(s^{2})
\end{eqnarray*}
For the carrying capacity, we will use a generalization of relation
(\ref{eq:CC}) :
\[
\frac{N^{2}}{\bar{N}^{2}}=(1-\delta+2\delta x)^{2}+2\epsilon\delta x(1-x)
\]
where $\bar{N}$ and $\delta$ were defined in (\ref{eq:Nbar},\ref{eq:delta}).
This relation reduces to (\ref{eq:CC}) when $\epsilon=0$ and ensures
that $N=\bar{N}=\mbox{const}$ when $\delta=0$. 

Repeating the computations of the previous section in the regime where
$\bar{N}s\ll1$ and $\delta\ll1$, and keeping the lowest order terms
leads to 
\[
\pi(x)=x-2\bar{N}sx(1-x)\left(1+\frac{2\epsilon}{3}(1-2x)\right)
\]
from which we can compute the ratio of invasion capacities $R_{SA}=\pi_{1}^{S}/\pi_{1}^{A}$
: 
\begin{eqnarray*}
R_{SA} & = & \left(\frac{1-\delta}{1+\delta}\right)\left(\frac{1+2\bar{N}s(1-2\epsilon/3)}{1-2\bar{N}s(1+2\epsilon/3)}\right)\\
 & \approx & 1+4\bar{N}s-2\delta
\end{eqnarray*}
The simplest diploid case (random mating, no linkage disequilibrium)
is similar to the haploid case and the criterion for cooperators to
prevail does not change.

\section{Conclusion.}

The problem of the emergence of cooperative behaviors and ``altruism''
has been a conundrum in evolutionary biology and has attracted a very
large number of contributions from different fields. We have shown
in this article that this conundrum may not exist at all, if we shift
our attention from deterministic advantages to fluctuation induced
advantages. The original Wright-Fisher model, developed in the 20's,
clarified the concepts of stochasticity in population genetics and
showed that a mutant, even when deleterious, has some probability
of invading the community, \emph{i.e. }$\pi_{1}^{A}>0$. We have,
in this article, extended this concept by showing that it is even
possible for the deleterious mutant to have a higher invasion capacity
than the wild type, \emph{i.e. }$\pi_{1}^{A}>\pi_{1}^{S}$. This is
based on the fact that purely fluctuation induced advantages can overcome
the disadvantages and the cost of cooperative behaviors if the relative
increase in the carrying capacity of the habitat induced by cooperators
is higher than the cost of altruism. This demonstration has been achieved
by the use of the generic Wright-Fisher model, which captures in very
simple terms the combined effects of finite size populations and deterministic
advantages.

All the existing models of cooperation (kin/multilevel/reciprocity/...)
have also been extended to finite populations in order to take into
account the importance of fluctuations. The important point to stress
is that in these models, there are always deterministic advantages
associated with cooperation. In other terms, the deterministic drift
term $a(x)$ has multiple zeros and thus the deterministic equation
for the proportion of $A$, 
\[
dx/dt=a(x)
\]
has more than one stable point, one of which corresponds to the dominance
of $A$ types. For example, the replicator dynamics used in the context
of evolutionary game theories uses\citep{Taylor2004}
\[
a(x)=x(1-x)\left(Ax+B(1-x)\right)
\]
and for $A>0,$ $B<0$ and $B/(B-A)\in[0,1]$, the dynamics possesses
two stable points $x=0$ and $x=1$. Taking into account fluctuations
and finite size populations then helps to explain how a single $A$
mutant can emerge and dominate the habitat\citep{Nowak2004}. The
same kind of arguments can be made for multilevel selection theories,
where the deterministic models already explain the possibility for
the existence of cooperators\citep{Silva1999} and then the computation
can be extended to take into account fluctuations\citep{Silva1999a}. 

As we stressed above, in the model we present in this article, there
is no deterministic advantages associated with cooperations and $a(x)<0$
for all $x\in[0,1]$. The only driving force in the present model
is provided by fluctuations due to finite size populations. The key
point, which we have demonstrated in the preceding sections, is that
the invasion capacity $\pi_{1}^{A}$ of the cooperators can be higher
than the invasion capacity of defectors $\pi_{1}^{S}$, even when
the cost of cooperation $s$ is \emph{always} positive and the deterministic
approach leads to the extinction of the $A$ types.

The aim of this article is not to contest the merits of existing models
such as kin or multilevel selection, which have been investigated
during the last forty years with a large number of case studies. I
propose an alternative approach to the problem of cooperation, that
is complimentary to the existing models and which restores the key
ingredients of population genetics to this field.

\paragraph{Acknowledgment.}

I am grateful to M. Vallade, E. Geissler and O. Rivoire for careful
reading of the manuscript and fruitful discussions. The author declares
to have no conflict of interest. 

\bibliographystyle{unsrt}
\bibliography{/home/bahram/0Papers/Bibtex/Altruism}

\end{document}